# Exploring co-sputtering of ZnO:Al and SiO$_2$ for efficient electron-selective contacts on silicon solar cells

Sihua Zhong[1,2,*], Monica Morales-Masis[1,3], Mathias Mews[4], Lars Korte[4], Quentin Jeangros[1], Weiliang Wu[1], Mathieu Boccard[1], and Christophe Ballif[1]

1 École Polytechnique Fédérale de Lausanne (EPFL), Institute of Microengineering (IMT), Photovoltaics and Thin-Film Electronics Laboratory (PV-lab), Rue de la Maladière 71b, CH-2002 Neuchâtel, Switzerland.
2. Shanghai Jiao Tong University, School of Physics and Astronomy, Institute of Solar Energy and Key Laboratory of Artificial Structures and Quantum Control (Ministry of Education), Shanghai 200240, P. R. China.
3. University of Twente, MESA+ Institute for Nanotechnology, 7500 AE Enschede, The Netherlands.
4. Helmholtz-Zentrum Berlin, Institute of Silicon Photovoltaics, Kekuléstraβe 5, D-12489 Berlin, Germany.

* Corresponding author: sihua.zhong@epfl.ch

## Abstract

In recent years, considerable efforts have been devoted to developing novel electron-selective materials for crystalline Si (c-Si) solar cells with the attempts to simplify the fabrication process and improve efficiency. In this study, ZnO:Al (AZO) is co-sputtered with SiO$_2$ to form AZO:SiO$_2$ films with different SiO$_2$ content. These nanometer-scale films, deposited on top of thin intrinsic hydrogenated amorphous silicon films and capped with low-work-function metal (such as Al and Mg), are demonstrated to function effectively as electron-selective contacts in c-Si solar cells. On the one hand, AZO:SiO$_2$ plays an important role in such electron-selective contact and its thickness is a critical parameter, thickness of 2 nm showing the best. On the other hand, at the optimal thickness of AZO:SiO$_2$, the open circuit voltage ($V_{OC}$) of the solar cells is found to be relatively insensitive to either the work function or the band gap of AZO:SiO$_2$. Whereas, regarding the fill factor ($FF$), AZO without SiO$_2$ content exhibits to be the optimal choice. By using AZO/Al as electron-selective contact, we successfully realize a 19.5%-efficient solar cell with $V_{OC}$ over 700 mV and $FF$ around 75%, which is the best result among c-Si solar cells using ZnO as electron-selective contact. Also, this work implies that efficient carrier-selective film can be made by magnetron sputtering method.

## 1. Introduction

Carrier-selective contacts applied to crystalline silicon (c-Si) solar cells are attractive due to their high efficiency and simple fabrication process. For example, c-Si solar cells using p-type hydrogenated amorphous silicon (p a-Si:H) as hole-selective layer and n-type a-Si:H film as the electron-selective layer have already been shown to reach 25.1% efficiency [1], and even 26.6% by further adopting an interdigitated back contact structure [2]. Due to these successes, a growing number of companies now produce solar cells that feature doped a-Si:H films as carrier-selective contacts. In research, other novel carrier-selective contacts based on polymer [3], metal oxide [4–7], metal fluoride [8,9], metal nitride [10] films, *etc*. have attracted a significant interest as these contacts have the potential to further improve the cell performance by using more transparent or conducting layers, and to simplify the fabrication process. Until now, various materials have been demonstrated as effective electron-selective layers, including LiF$_x$ [8], MgF$_x$ [9], MgO$_x$ [6], TiO$_x$ [4,5], TaO$_x$ [11], TaN$_x$ [10], alkali/alkaline-earth metal carbonates [12], and their combinations [13], in some cases combined with intrinsic a-Si:H (i a-Si:H) for passivation.

ZnO (with or without doping), one of the most widely used transparent conductive oxides [14–16], has also been proposed as an electron-selective layer in c-Si solar cells [17–20] due to the fact that the conduction band offset between c-Si and ZnO is beneficial for electron transport but the valence band offset forms a barrier for holes transport from c-Si to ZnO [18]. Combining i a-Si:H and ZnO:B grown by metal-organic chemical vapor deposition as the electron-selective layer, an efficiency of 16.6% was demonstrated by Wang *et al* [18]. Ding *et al*. [20] realized an 18.46%-efficient c-Si solar cell by using spin-coated ZnO:Al (AZO) on top of i a-Si:H as the electron-selective layer. The researchers from Ye group also demonstrated spin-coated AZO as an effective electron-selective film, achieving an efficiency of 13.6% [19]. Recently, they doped ZnO with Li to reduce its work function and adding intrinsic a-Si:H as passivation layer, promoting the efficiency to 15.1% [21].

In this study, magnetron sputtering method is utilized to prepare AZO as electron-selective film, which is a method easy to fabricate uniform films on large-size substrates and highly compatible to the present mass-production line of solar cells. We co-sputter AZO (2 wt% $Al_2O_3$) and $SiO_2$ to form AZO:$SiO_2$ films, which is demonstrated to be a low-work-function material [22]. We investigate the influence of the power applied to the $SiO_2$ target ($P_{SiO2}$) and the $O_2$/(Ar+$O_2$) flow ratio on the microstructure, conductivity, work function and band gap of the AZO:$SiO_2$ films. When AZO:$SiO_2$ capped with a thermally evaporated metal are applied to c-Si solar cells as electron-selective contacts, it is find that both the thickness of AZO:$SiO_2$ and the work function of the capping metal are highly important. At the optimal thickness of AZO:$SiO_2$, we show how the deposition conditions of AZO:$SiO_2$ affects the open circuit voltage ($V_{OC}$) and fill factor ($FF$), and thus pure AZO, i.e. $P_{SiO2}$=0 W, is found to be the best choice. Finally, we realize a 19.5%-efficient c-Si solar cell by using an AZO/Al stack (on top of a passivating i a-Si:H layer) as electron-selective contact, which is the highest efficiency among the solar cells using ZnO as electron-selective film. In addition, it is worth mentioning that the previously successful electron-selective films are mainly made by either thermal evaporation [6,8,9,12] or atomic layer deposition [5,10,11]. Here our study firstly shows that magnetron sputtering is also a feasible method to fabricate efficient electron-selective films.

## 2. Results and discussion

Fig.1(a and b) compares the microstructure of AZO and AZO:$SiO_2$ films by plane-view transmission electron microscopy (TEM) images. The film thickness was ~2 nm in both cases. The power applied to the AZO target was 35 W, which was maintained for the entire study. For Fig.1(b), $P_{SiO2}$ was 25 W. Both AZO and $SiO_2$ targets had a diameter of 100 mm. Completely different microstructures of the AZO and AZO:$SiO_2$ are observed from the scanning TEM (STEM) images (Supporting information, Fig.S1). High resolution TEM images (Fig.1(a) and (b)) further reveal that the AZO film contains crystallites with a wurtzite structure and a diameter of a few nanometers in an amorphous matrix, whereas the AZO:$SiO_2$ film is fully amorphous. A gradual decrease in crystallinity with increased $SiO_2$ content could be observed in other studies [22,23]. The influence of the addition of $SiO_2$ on the optoelectronic properties of these layers is presented below.

We performed X-ray photoelectron spectroscopy (XPS) measurements to determine the ratio of Si to (Si+Zn) as a function of $P_{SiO2}$ and $O_2$/(Ar+$O_2$) flow ratio during sputtering. As shown in Fig.1(c), on the one hand, Si/(Si+Zn) increases from 0 to ~0.4 when increasing the $P_{SiO2}$ from 0 W to 35 W at a constant $O_2$/(Ar+$O_2$) flow ratio of 0.18%. On the other hand, Si/(Si+Zn) decreases with increasing the $O_2$/(Ar+$O_2$) at a constant $P_{SiO2}$ of 25 W, which indicates that adding $O_2$ during sputtering reduces the incorporation of $SiO_2$ into AZO:$SiO_2$.

Fig.1(d) shows the conductivity of the different AZO:$SiO_2$ films. Increasing $P_{SiO2}$ leads to a drastic decrease of the conductivity, which is linked to the higher Si/(Si+Zn) ratio in the film and its amorphization [22]. When $P_{SiO2}$ is 0 W, namely for a pure AZO film, the conductivity decreases with increasing $O_2$/(Ar+$O_2$) flow ratio, coinciding with the literature [24]. In other cases ($P_{SiO2}$ > 0 W), the film conductivity increases with $O_2$/(Ar+$O_2$) flow ratio, owing to the lower $SiO_x$ incorporation at higher $O_2$/(Ar+$O_2$).

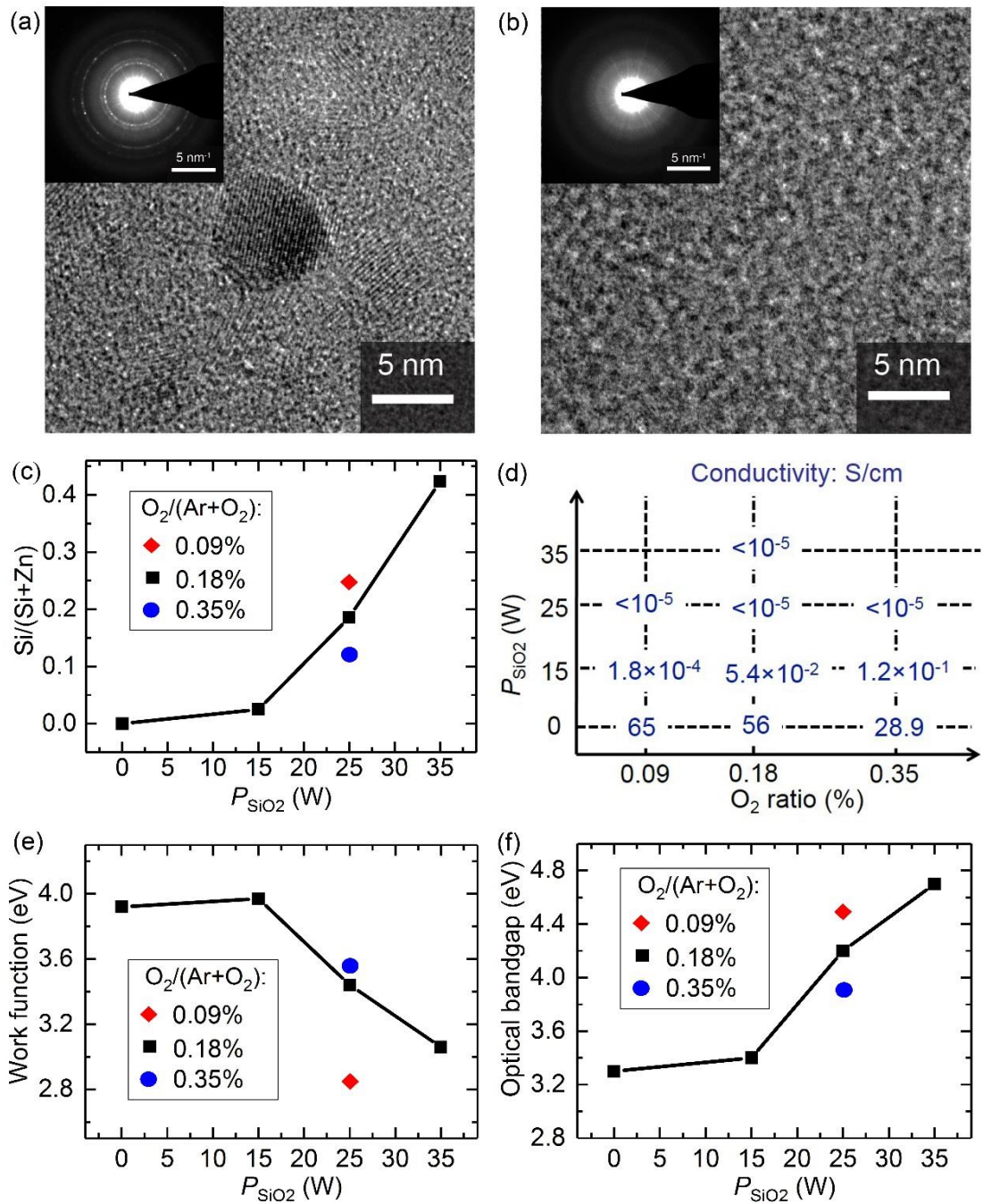

Fig.1 High-resolution plane-view TEM images of (a) AZO film deposited under $O_2/(Ar+O_2)$: 0.18%, and (b) AZO:SiO$_2$ film deposited under $P_{SiO2}$: 25 W, $O_2/(Ar+O_2)$: 0.18%. Insets show the selected-area electron diffraction patterns. (c) Si/(Si+Zn) ratios (from XPS), (d) conductivity, (e) work function and (f) optical band gap of the AZO:SiO$_2$ film versus the power applied to the SiO$_2$ target ($P_{SiO2}$) and for different $O_2$ to (Ar+O$_2$) flow ratio in the sputtering gas.

The work function of the AZO:SiO$_2$ films was also investigated by Helium ultra-violet photoelectron spectroscopy (He-UPS). The results are shown in Fig.1(e). Increasing the $P_{SiO2}$ and hence the Si/(Si+Zn) ratio results in lower work functions, which agrees well with the result of Nakamura *et al* [22]. Furthermore, reducing the $O_2/(Ar+O_2)$ flow ratio also leads to a lower work function, owing to the more efficient SiO$_2$ incorporation. To further study the influence of the deposition conditions ($P_{SiO2}$ and $O_2/(Ar+O_2)$ flow ratio) on the bandgap of AZO:SiO$_2$ films, optical absorption coefficients were determined using UV-Vis-NIR spectroscopy. Through the Tauc plots assuming that the films have direct bandgap [25] (see Fig.S2, supporting information), the optical bandgap is obtained. As shown in Fig.1(f), both increasing the $P_{SiO2}$ and reducing the $O_2/(Ar+O_2)$ flow ratio lead to higher bandgap.

In summary, variations of $P_{SiO2}$ and $O_2/(Ar+O_2)$ flow ratio in this study yield significant changes to the material properties. These changes are expected to affect the solar cell performance when using this material as electron-selective contact, which is discussed in the following.

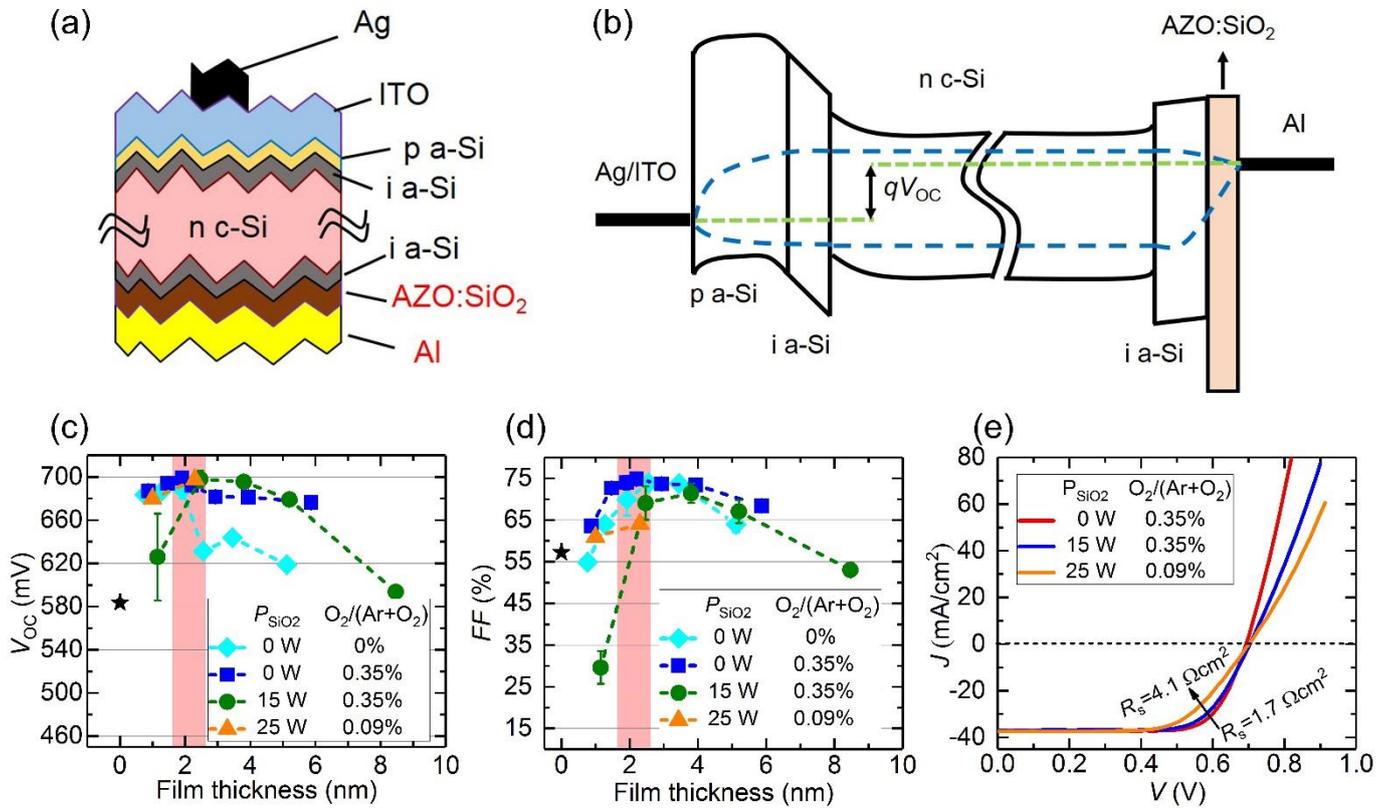

Fig.2 Schematic cross section (a) and band diagram at open-circuit conditions (b) of the solar cells using AZO:SiO$_2$/Al as electron-selective contacts. (c) $V_{OC}$ and (d) $FF$ of the solar cells with AZO:SiO$_2$ films deposited at different conditions (different $P_{SiO2}$ and $O_2/(Ar+O_2)$ ratio). Average of 5 cells is displayed and error bar displays the standard deviation. (e) $J$-$V$ curves of the solar cells using different (~2-nm-thick) AZO:SiO$_2$/Al as electron-selective contacts.

Fig.2(a) and (b) show the structure and a schematic band diagram of solar cells using AZO:SiO$_2$/Al as electron-selective contact stack and p-type a-Si:H as hole selective contact. Intrinsic a-Si:H was used on both sides of the n-type c-Si wafer as passivation layer. For the AZO:SiO$_2$ films, different thicknesses, different $P_{SiO2}$ and $O_2/(Ar+O_2)$ flow ratios have been studied. In this section, AZO:SiO$_2$ includes pure AZO (*i.e.* films prepared with $P_{SiO2} = 0$ W). Fig.2(c) shows that the $V_{OC}$ is only around 580 mV if Al directly covers on i a-Si:H film. However, when a thin (1 nm-thick) AZO:SiO$_2$ layer is inserted between the i a-Si:H and Al, $V_{OC}$ is greatly increased, demonstrating the significance of AZO:SiO$_2$ on the electron-selective contact. The presence of AZO:SiO$_2$ can remove the Fermi-level pinning between Al and i a-Si:H and may also reduce the carrier recombination at the interface.

For a thickness of AZO:SiO$_2$ around 2 nm, best $V_{OC}$ values around 690-700 mV are obtained. Interestingly, despite the different deposition conditions ($P_{SiO2}$ and $O_2/(Ar+O_2)$) and thus the different material properties as discussed in previous section, very similar $V_{OC}$ values can be reached. Additionally, very similar $V_{OC}$ values can also be obtained at a thickness of ~2 nm by using undoped ZnO capped with Al as electron-selective contact stack (see Supporting information, Fig. S3). Since $V_{OC}$ is determined by the difference between hole quasi-Fermi level at the positive electrode and electron quasi-Fermi level at the negative electrode, as shown in Fig.2(b), and based on the fact that i a-Si:H films capped with different AZO:SiO$_2$ have the same passivation quality (they have comparable implied $V_{OC}$ within the range of 736-741 mV), thus the measurement of very similar $V_{OC}$ values implies that electron quasi-Fermi levels at the negative contact are almost the same for the solar cells with different AZO: $SiO_2$ films. We hypothesize that this is because the AZO:$SiO_2$ films are too thin to screen the influence of Al, making the band diagram dominated by the Al properties and not by the AZO:$SiO_2$ properties. Simulations were also carried out to observe the insensitivity of $V_{OC}$ to the work function of an ultrathin film, as shown in Fig.S4 of Supporting information. The results may help explain that other ultrathin films such as $LiF_x$ [8], $MgF_x$ [9], $MgO_x$ [6], $SiO_2$ [26], *etc*., combined with a low work function metal can work similarly well as electron-selective contact despite their different energy-band structures. Nevertheless, this hypothesis does not imply that any material can work well as electron-selective contact, since different materials may have different abilities to screen the metal work function due to some different material properties, e.g. effective conduction band density of states (see Supporting information, Fig.S4). Also, the way in which the material affects the effective work function of metal is important.

Fig.2(d) further shows that similarly to Voc, *FF* increases first and then decreases with increasing the thickness of AZO:$SiO_2$ films. Note that for 0 nm of AZO:$SiO_2$, the *FF* is variable from run to run. For solar cells made with an AZO:$SiO_2$ thickness of around 2 nm, *FF* decreases with increasing $P_{SiO2}$. The current density-voltage (*J-V*) curves under air mass 1.5 global (AM1.5G) illumination of cells made with a 2-nm-thick AZO:$SiO_2$ layer are shown in Fig.2(e), from which the influence of series resistance ($R_S$) on *J-V* curves is observed. Based on the method proposed by Bowden [27], $R_S$ is calculated to increase from 1.7 $\Omega cm^2$ for cells with pure AZO to 4.1 $\Omega cm^2$ for cells with AZO:$SiO_2$ (25 W $SiO_2$), which results in the reduced *FF*. The increased $R_S$ is correlated with the decreased conductivity of the AZO:$SiO_2$ film with increasing $P_{SiO2}$, as presented in Fig 1(d). Therefore, although the AZO:$SiO_2$ deposition conditions have no influence on $V_{OC}$, they do affect *FF*.

To get further insights into the working mechanisms of AZO:$SiO_2$/metal electron-selective contact stacks, different capping metals have been studied for different AZO:$SiO_2$ film thickness. Here, the $P_{SiO2}$ was set to 15 W and the $O_2/(Ar+O_2)$ flow ratio to 0.18%. The solar cells maintain the same structure as shown in Fig.2(a), but Mg, Al, Cu and Au are used as the negative electrodes. Based on literature data, their bulk work functions are 3.66 eV, 4.06-4.26 eV, 4.48-5.1 eV, 5.31-5.47 eV, respectively [28]. However, the effective work function of metals can change depending on the deposition conditions and the formation of the interface to the film that is contacted (e.g., formation of interface dipoles, Fermi-level pinning). Fig.3(a) schematically shows the possible energy band diagram near negative contact of the solar cells with the ultrathin AZO:$SiO_2$ film/metal as electron-selective contact under open-circuit conditions. Note, that band bending in the AZO:$SiO_2$ layers is not represented (although it is expected to be significant), and that the metal work function does not precisely correspond to the literature value, which will be explained in the following. As presented, higher effective work function of the metal leads to upwards band bending in the silicon wafer, i a-Si:H layer and AZO:$SiO_2$. This reduces selectivity of the contact, leading to a slope of electron quasi-Fermi level in the i a-Si:H and AZO:$SiO_2$ layers. The band bending in the silicon wafer also results in an increase in hole concentration at the electron contact, increasing carrier recombination, thus leading to smaller Fermi-level splitting in the absorber. Due to these reasons, the $V_{OC}$ is expected to be lower, which is confirmed in Fig.3(b). In addition, a higher upwards band bending means a higher energy barrier for electrons to be collected and thus lower *FF*, as verified in Fig.3(c). Similar result is also reported in the solar cells using $TiO_x$ capped with metal as carrier-selective contact [29].

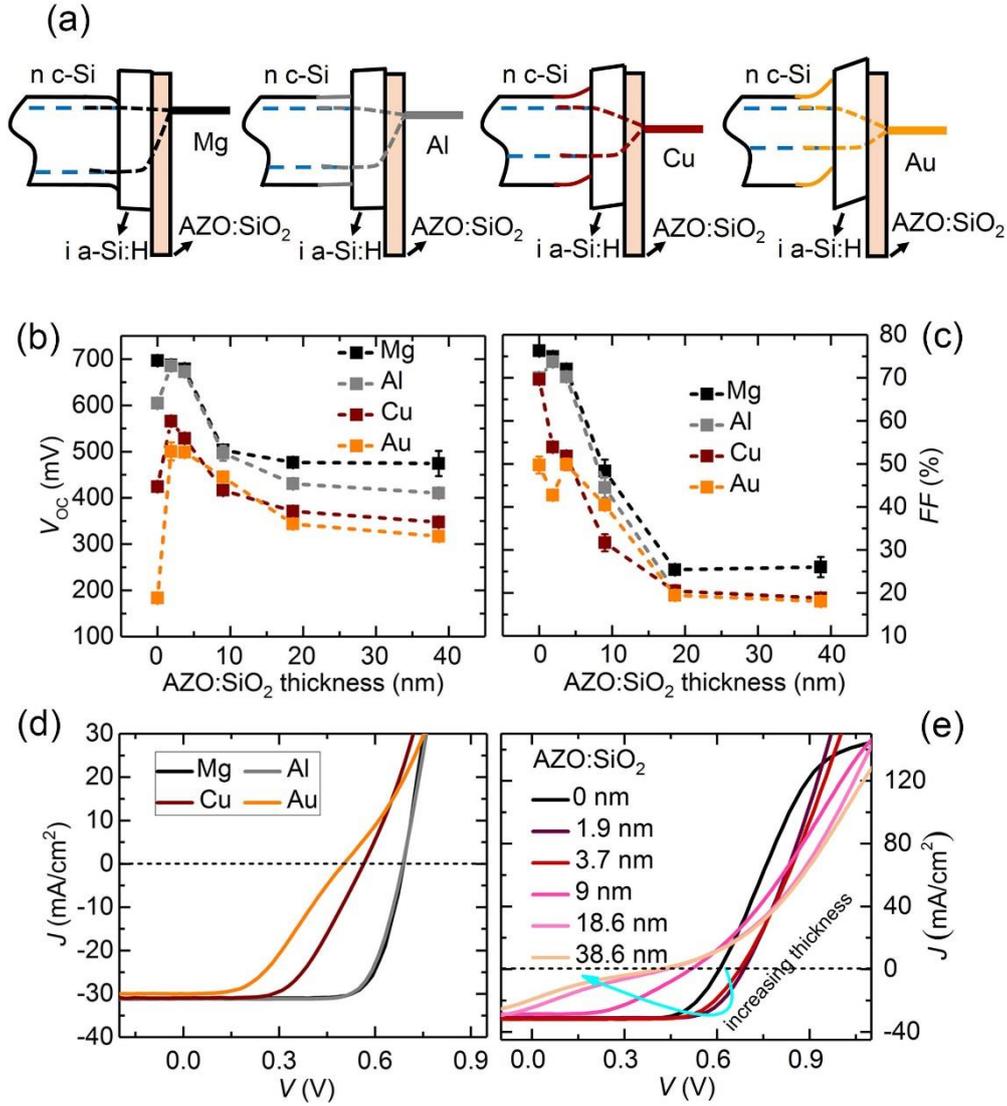

Fig.3 (a) Energy band diagram near negative contact of the solar cells using ultrathin AZO:SiO$_2$ film and different capping metals as electron-selective contact stack at open-circuit conditions. Energy band bending in the AZO:SiO$_2$ is neglected. (b) $V_{OC}$ and (c) $FF$ varying with the metal and the thickness of the AZO:SiO$_2$ film. Average of 3 cells is displayed and error bar displays the standard deviation. (d) $J$-$V$ curves of the solar cells with 2-nm-thick AZO:SiO$_2$ films but different capping metals. (e) $J$-$V$ curves of the solar cells varying with the thickness of AZO:SiO$_2$ films. The capping metal is Al.

When in the absence of any AZO:SiO$_2$ film, athough both $V_{OC}$ and $FF$ obviously depend on the metal, the pinning factor of Fermi-level between the metal layers and the i a-Si:H film is estimated to be 0.3 according to the method described in literature [30]. When the thickness of the AZO:SiO$_2$ film increases to ~2 nm, the Fermi-level pinning between the metal layers and the i a-Si:H film is removed but the pinning factor of Fermi-level between the metal and AZO:SiO$_2$ is estimated to be 0.1, showing more severe Fermi-level pinning effect. However, from the significant improvement of the $V_{OC}$s of solar cells with Al, Cu and Au, it can be speculated that the pinning position of Fermi-level should move to a higher energy level. It is worth mentioning that inserting TiO$_2$ between c-Si and metal is also reported to change the Fermi-level pining position [31]. The shifting of the Fermi-level pining position modifies the effective work function of metal, a reduction for Al, Cu and Au, but an increase for Mg. This makes the effective work function of Mg similar to that of Al, but lower than that of Cu and Au. Hence the $V_{OC}$s of solar cells with Al and Mg are similar but higher than that of cells with Cu and Au when the thickness of the AZO:SiO$_2$ is ~2 nm.

Fig.3(d) further shows the $J$-$V$ curves of the solar cells using different metals but with the same

2-nm-thick AZO:SiO$_2$ film under AM1.5G illumination. The *J-V* curve of the solar cell with Au obviously deviates from that of a diode, suggesting that there is a strong n c-Si/Au Schottky diode opposing the solar cell diode.

For AZO:SiO$_2$ film thicknesses between 2 nm and 20 nm, both the $V_{OC}$ and *FF* decrease with increasing AZO:SiO$_2$ thickness for any capping metal. Note that this decrease is not caused by the increased sputtering damage with time since sputtering damage is almost eliminated thanks to the low sputtering power and appropriate annealing process (see Supporting information, Fig.S5). For AZO:SiO$_2$ thicknesses above 20 nm, both $V_{OC}$ and *FF* become insensitive to the thickness and appear to stabilize to poor values.

Fig.3(e) shows the influence of the thickness of AZO:SiO$_2$ film on the *J-V* curves of the Al contacted solar cell under AM1.5G illumination. With no AZO:SiO$_2$, the *J-V* curve shows an S shape, probably because of a Schottky contact between Al and the n c-Si wafer passivated with intrinsic a-Si:H. When the AZO:SiO$_2$ thickness increases to ~2 nm, the effective work function of Al is reduced as mentioned above, and a diode *J-V* curve is obtained. Further increasing the thickness, the *J-V* curves are influenced by the increased resistance and deviate from that of a single diode, which lead to the decrease of *FF*.

Finally, to show the potential of sputtered AZO:SiO$_2$ as electron-selective contact, a 4-cm$^2$ cell was made using a 2-nm-thick simple AZO film capped with Al as electron-selective contact. A 19.5% efficiency is demonstrated with a $V_{OC}$ of 701 mV and a *FF* of 74.7%. The *J-V* curve of this device is shown in Fig.4. This is the highest efficiency reported for a solar cell that features ZnO as electron-selective contact. This result is comparable to those of cells using other successfully demonstrated electron-selective films [8–12] and also the result suggests that magnetron sputtering method can be used to make efficient electron-selective films, which is one of the methods easy applied to mass production line. Nevertheless, it is necessary to point out that the *FF* is still limited by relatively high series resistance, even for this optimal composition and thickness. Also, a striking and universal problem for using nanometer-thin films combined with low work function metal as carrier-selective contact is the relatively low $J_{SC}$. One of the reasons is infrared absorption losses owing to metal close to the Si wafer [32]. We believe that utilizing a thick and conductive film as electron-selective layer is required to avoid this infrared absorption and fully benefit from the novel carrier-selective contact approach.

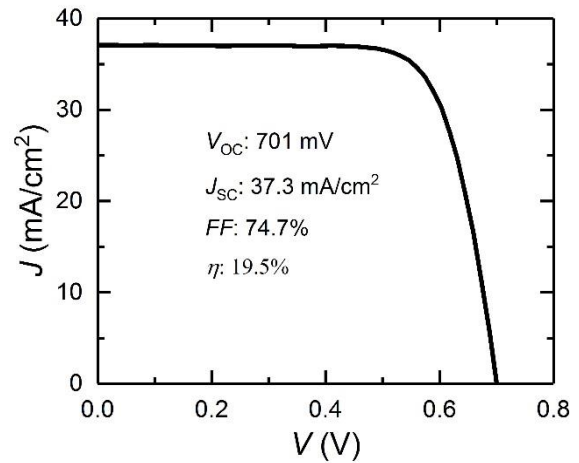

*Fig.4 J-V curve of the best solar cell using AZO/Al as electron-selective contact.*

## 3. Conclusion

In summary, ZnO:Al (AZO) is co-sputtered with SiO$_2$ to form AZO:SiO$_2$ films with different SiO$_2$ content. These films, capped with different metals, have been applied as electron-selective contact in c-Si solar cells. The microstructure of the AZO film can be changed by incorporating SiO$_2$. Both increasing the power applied to the SiO$_2$ target and decreasing O$_2$/(Ar+O$_2$) flow ratio lead to higher Si/(Si+Zn), resulting in lower conductivity, lower work function and enlarged bandgap. On the one hand, thickness of these films was shown to be a critical parameter when applying them as electron-selective contacts, a thickness of ~2 nm appearing as optimal. At this thickness, the performance of solar cells is significantly improved compared to that without AZO:SiO$_2$ film. On the other hand, the open circuit voltage ($V_{OC}$) was found to be insensitive to the deposition conditions of AZO:SiO$_2$, despite the variation of work function and bandgap of the material. However, the deposition condition of the AZO:SiO$_2$ film was shown to greatly affect fill factor (*FF*). AZO, without SiO$_2$ content, was thus shown to be the best film. Furthermore, we showed that the effective work function of the capping metal has a significant influence on both $V_{OC}$ and $FF$, much more striking than the AZO:SiO2 material properties. Finally, a 19.5%-efficient c-Si solar cell with $V_{OC}$ of 701 mV and $FF$ of 74.7% is demonstrated by using AZO/Al as electron-selective contact. This study successfully shows that magnetron sputtering is capable of making efficient carrier-selective films. Further improvements will rely largely on improving the optical properties of this system, e.g. by inserting a >100-nm-thick low-refractive-index optical spacer between the wafer and the metal.

## 4. Experimental section

The AZO:SiO$_2$ films were deposited by RF-co-sputtering AZO (Al$_2$O$_3$: 2 wt%) and SiO$_2$ at room temperature in a magnetron sputtering system (Leybold Univex). The target diameter was 100 mm. The sputtering power of AZO was fixed at 35 W, and the $P_{SiO2}$ varied from 0 W to 35 W. The flow rates of Ar and O$_2$ were changed to yield an O$_2$/(Ar+ O$_2$) flow ratio of 0%-0.35%, the working pressure was fixed at $2.7\times10^{-3}$ mbar. The substrate was rotated at 10 rpm to obtain homogeneous films. Film thickness was controlled by the deposition time.

For solar-cell fabrication, n-type float zone (FZ) c-Si wafers with resistivity of 1-5 Ω cm and thickness of either 240 μm (for Fig.2 and 4 ) or 180 μm (for Fig.3) were used. These wafers were firstly etched in KOH solution to form randomly pyramids-textured surface, followed by wet-chemical cleaning and HF dipping. Intrinsic a-Si:H films were then deposited on both sides of the Si wafers as passivation layer by plasma enhanced chemical vapor deposition (PECVD). p-type a-Si:H was deposited on the front side (*i.e.* illumination side) of the wafer, and the AZO:SiO$_2$ film was deposited on the back side by magnetron sputtering. The front side was then covered with an ~80-nm-thick indium-tin-oxide (ITO) film as the antireflection and conductive layer, and the front metal grids were prepared by either magnetron sputtering through a shadow mask for the 1.1-cm$^2$ cells (Fig.3) or screen printing for the 4-cm$^2$ cells (Fig.2 and 4), followed by annealing at 210 °C for 30 minutes. Finally, the back side of the wafer was covered with the metal film by thermal evaporation.

Characterization: The Si to (Si+Zn) ratio of the AZO:SiO$_2$ films were characterized by X-ray photoelectron spectroscopy (XPS) with Al-K$_α$ excitation. To this end the Si 2p, O 1s and Zn 3p core levels were measured and fitted using a linear background and Voigt peaks with a 15 % Lorentz-contribution. The Zn 3p peak was fitted using two peaks with a fixed distance of 2.95 eV and the same full-width at half maximum. These two peaks represent the contributions from ZnO and ZnOH. The Si 2p signal was fitted with a single signal and the O 1s signal was fitted with two signals to account for SiO and ZnO contributions. The Si and Zn contents of the mixed layers were calculated using sensitivity factors, extracted from stochiometric ZnO and SiO2 samples. The ratio of the Si/Zn oxide peak area to the O 1s peak area, corrected by the stoichiometry of the respective element, was used as the sensitivity factor. These sensitivity factors were used to obtain the fraction of Si and Zn in the in the mixture. The work function of the films was characterized using Helium ultra-violet photoelectron spectroscopy (He-UPS). A bias voltage of 5 eV was applied and the secondary electron cut-off was measured and fitted using a Boltzmann-Sigmoid function to obtain the work function of the layers.

The TEM observation of the AZO:SiO$_2$ films were performed using an FEI Tecnai Osiris microscope. For that purpose, AZO and AZO:SiO$_2$ thin films were directly sputtered onto Cu grids coating with a thin C film. High-resolution TEM top view images were recorded alongside selected-area electron diffraction patterns to assess the microstructure of the films. The reflectance spectra and transmittance spectra of samples were measured with a spectrophotometer (Lambda-950, Perkin Elmer) to extract the absorption coefficients. The thickness of the AZO:SiO$_2$ film on planar surface was measured by ellipsometry, and a factor 0.66 was applied to obtain an estimate of that on the textured Si surface. Al electrodes with 1-mm spacing were deposited on the AZO:SiO$_2$ film to measure the dark conductivity (The measured value is the lower limit of the real conductivity of the films since the contact resistance is included). Solar cell characterizations were carried out using a Wacom WXS-90S-L2 solar simulator, at standard test conditions (AM1.5G spectrum, 100 mW/cm$^2$ and 25 °C).

# Acknowledgements

The authors thank Raphaël Monnard for amorphous silicon preparation, and thank Christophe Allebe, Fabien Debrot and Nicolas Badel from CSEM for the high quality wet-processing and metallization. This project has received funding from the European Union's Horizon 2020 research and innovation programme under Grant Agreement No. 727523 (NextBase), as well as Swiss national science foundation under Ambizione Energy grant ICONS and the China Postdoctoral Science Foundation.

# Notes

The authors declare no competing financial interest.

# Supporting information

Supporting information associated with this article can be found in the online version.